\documentstyle[12pt,epsf]{article}
\newcommand{\beq}[1]{\begin{equation}\label{#1}}
\newcommand{\eeq}{\end{equation}}
\newcommand{\beqar}{\begin{eqnarray}}
\newcommand{\eeqar}{\end{eqnarray}}
\newcommand{\ttq}{$t \bar t$}
\newcommand{\pb}{{\bf p}}
\newcommand{\kb}{{\bf k}}

\def\app#1#2#3{{\it Act. Phys. Pol. }{\bf B #1} (#2) #3}

\def\npb#1#2#3{{\it Nucl. Phys. }{\bf B #1} (#2) #3}
\def\plb#1#2#3{{\it Phys. Lett. }{\bf B #1} (#2) #3}
\def\prd#1#2#3{{\it Phys. Rev. }{\bf D #1} (#2) #3}

\def\prl#1#2#3{{\it Phys. Rev. Lett. }{\bf #1} (#2) #3}

\def\sovnp#1#2#3{{\it Sov. J. Nucl. Phys. }{\bf #1} (#2) #3}

\def\zpc#1#2#3{{\it Z. Phys. }{\bf C #1} (#2) #3}

\begin{document}
\thispagestyle{empty}
\begin{titlepage}
\hskip -3cm
\begin{flushright}
{\bf TTP 95--25\footnote{The complete postscript file of this preprint,
	including figures, is
	also available via anonymous ftp as /ttp95-25/ttp95-25.ps at
	ttpux2.physik.uni-karlsruhe.de (129.13.102.139), or via www at
	http://ttpux2.physik.uni-karlsruhe.de/cgi-bin/preprints/.}}\\
{\bf May 1995}\\
{\bf hep-ph/9506292}
\end{flushright}
\vspace{1cm}
\begin{center}
{\Large\bf Higgs effects in top quark pair production\footnote{
	Work supported in part by KBN grant 2P30225206 and by EEC network
	CIPDCT 940016.}}
\end{center}
\vspace{0.8cm}
\begin{center}
{\sc R. Harlander$^a$, M. Je\.zabek$^{a,b}$ and J.H. K\"uhn$^{a,c}$}
\vspace*{2mm} \\
{\sl 	$^a$ Institut f\"{u}r Theoretische Teilchenphysik,
     	D--76128 Karlsruhe, Germany \\
     	$^b$ Institute of Nuclear Physics, Kawiory 26a, PL--30055 Cracow,
	Poland\\
	$^c$    Stanford Linear Accelerator Center, Stanford, CA 94309}
\end{center}
\vspace{2.cm}
\begin{abstract}
\noindent
Top quark production in $p\bar p$ and $e^+e^-$ collisions is enhanced by the
exchange of a Higgs boson. The enhancement factors are calculated in the
threshold region using the Greens function method.
\end{abstract}
\end{titlepage}
\setcounter{page}{2}
The recent discovery of the top quark \cite{CDF,dnull} and its large mass
following
from both direct \cite{CDF,dnull} and indirect \cite{schaile} observations
necessitate evaluation of Higgs effects for top quark pair production
in the threshold region\footnote{
	The top mass measured by the CDF Collaboration is
	$m_t = 176\pm 8 \mbox{(stat.)} \pm 10 \mbox{(sys.)}$ GeV
	and the value obtained by the D$0\!\!\!/$ Collaboration is equal to
	$m_t = 199^{+19}_{-21} \mbox{(stat.)} \pm 22 \mbox{(sys.)}$ GeV.
	Indirect determination \cite{schaile} gives
	$m_t = 178 \pm 11 ^{+18}_{-19}$ GeV.}.
It has been observed long ago \cite{fadkhoz,fks,strassler} that these effects
can be important for heavy top quarks due to the attractive Yukawa force
in \ttq--systems. Although this force is weaker than chromostatic
interactions its quantitative description will be necessary in future precise
studies of the threshold top pair production at both $e^+e^-$ and $p\bar p$
colliders.
In this article enhancement factors are evaluated for both singlet and octet
color configuration of the \ttq--system. The former is relevant for
$e^+e^-$ annihilation and the latter for $q\bar q$ annihilation which
dominates at TEVATRON energies. In both cases the effects of Higgs exchange
enhance the production cross sections.
The perturbative potential depends on the color configuration of \ttq.
It is proportional to the color factor $C_R$. For the singlet configuration
$C_R=C_1 = -4/3$, for  octet $C_R=C_8 = +1/6$.
Thus the chromostatic interactions
are attractive for singlet and repulsive for octet configurations\footnote{
	The relation $C_1+8 C_8 = 0$ follows from the tracelessness of
	the Gell--Mann matrices:
	${\rm Tr}\sum_{a=1}^8 \lambda_a^{(1)} \lambda_a^{(2)} =0$.
	The latter equation means that the sum of the color factors for single
	gluon exchange in the \ttq--system vanishes after averaging over
	all color configurations of \ttq. Thus the sum of the color factors
	for one singlet and eight octet states is equal to zero.}.
The long distance part of the chromostatic potential, which is presumably
related to some composite scalar exchange, may depend in a different way
on the color configuration of \ttq. However, we have checked that the short
lifetime of top quarks cuts off contributions corresponding to small momentum
transfers.

The threshold behavior of \ttq--production is determined
by the $s$-- and $p$--wave Greens functions $G(\pb,E)$ and $F(\pb,E)$
satisfying the following Lippmann--Schwinger equations:
\beqar
G(\pb,E) &=& G_0(\pb,E) + G_0(\pb,E) \int {d^3 k\over (2 \pi)^3}
        V(\pb - \kb) G(\kb,E) 			\label{lipps} \\
F(\pb,E) &=& G_0(\pb,E) + G_0(\pb,E) \int {d^3 k\over (2 \pi)^3}
        {\pb \cdot \kb \over \pb^2}
        V(\pb - \kb) F(\kb,E) \ \ ,  		\label{lippp}
\eeqar
where
\beq{prop_free}
G_0(\pb,E) = {1\over E - {\pb^2 \over m_t} + i \Gamma_t}
\eeq
and $E = \sqrt s - 2 m_t$.
$\Gamma_t$ is the width of the top quark taken to be constant.

In \cite{JKT,hjkt} these equations were solved numerically for a static
potential
\beq{v_qcd}
V_{\rm QCD}(\pb) = C_R {\alpha_{eff} (\pb) \over \pb^2} \ \ ,
\eeq
assuming instantaneous gluon exchange between the two quarks.
$C_R$ is an $SU(3)$--group theory factor depending on the color state
of the \ttq--system.
The function $\alpha_{eff}$ is described in \cite{JT}. Its form follows
from the assumption that the QCD--potential is described by
a phenomenological Richardson ansatz for small momentum transfers and the
perturbative 2--loop formula for intermediate and large ones.

To take Higgs effects into account we follow \cite{jezk} using
the Yukawa potential approach combined with perturbative results.
This means that we insert the effective potential
\beq{v_eff}
V_{\rm eff} = V_{\rm QCD} + V_{\rm Yuk}
\eeq
into eqs.(\ref{lipps}) and (\ref{lippp}), with $V_{\rm QCD}$ from (\ref{v_qcd})
and
\beq{v_yuk}
V_{\rm Yuk}(\pb) = -{4 \pi \kappa \over m_H^2 + \pb^2} \ \ ,
\eeq
\beq{kappa}
\kappa = \sqrt 2 G_F m_t^2 / 4\pi \ \ .
\eeq
The outcome $G(\pb,E)$ for the numerical evaluation of (\ref{lipps})
then has to be multiplied by the energy independent factor
\beq{corfac}
1+{\kappa \over \pi} \widehat{f}_V(m_H^2/m_t^2) \ \ ,
\eeq
with
\beq{fhat}
\widehat{f}_V(m_H^2/m_t^2) = f_{\rm thr}(m_H^2/m_t^2) -
	\pi {m_t \over m_H} \ \ .
\eeq
$f_{\rm thr}$ is the perturbatively evaluated threshold correction function
and can be written as \cite{guth}
\beq{fthr}
f_{\rm thr}(r) = -{1 \over 12} \bigg[ -12 + 4 r + (-12 + 9 r - 2 r^2)\, \ln r
	+ {2 \over r} ( -6 + 5 r - 2 r^2) l_4(r)\bigg] \ ,
\eeq
with
\beq{l4}
l_4(r) = \left\{ \begin{array}{l@{\quad\mbox{if}\quad}l}
		\sqrt{r(4-r)} \arccos (\sqrt{r}/2) & r \le 4 \ , \\
		-\sqrt{r (r-4)} {1 \over 2} \,
		\ln {1+\sqrt{1-4/r} \over 1-\sqrt{1-4/r}} & r > 4 \ .
		\end{array}
	\right.
\eeq

For the singlet case the effect on the total cross section can therefore
be summarized by
the following replacement of the lowest order phase space factor
$\beta = \sqrt{1-4 m_t^2/s}$:
\beq{phase_1}
\beta \rightarrow \left( 1-{8 \alpha_s \over 3 \pi} \right)^2 B(E) \ \ ,
\eeq
with
\beq{phase_sp}
B(E) = \left( 1+{\kappa \over \pi} \widehat{f}_V ({m_H^2/m_t^2})
		\right)^2
	{2 \Gamma_t \over m_t^2 \pi} \int_0^\infty dp\, p^2 |G(\pb,E)|^2 \ \ .
\eeq

Expecting the main effect to come from the modification
$V_{\rm QCD} \rightarrow V_{\rm QCD} + V_{\rm Yuk}$ and realizing
the attractive character of $V_{\rm Yuk}$,
the production cross section should be enhanced with the
global factor (\ref{corfac}) giving damping corrections to this enhancement.

\begin{figure}
\begin{center}
\leavevmode
\epsfxsize=11.5cm
\epsffile[90 120 470 700]{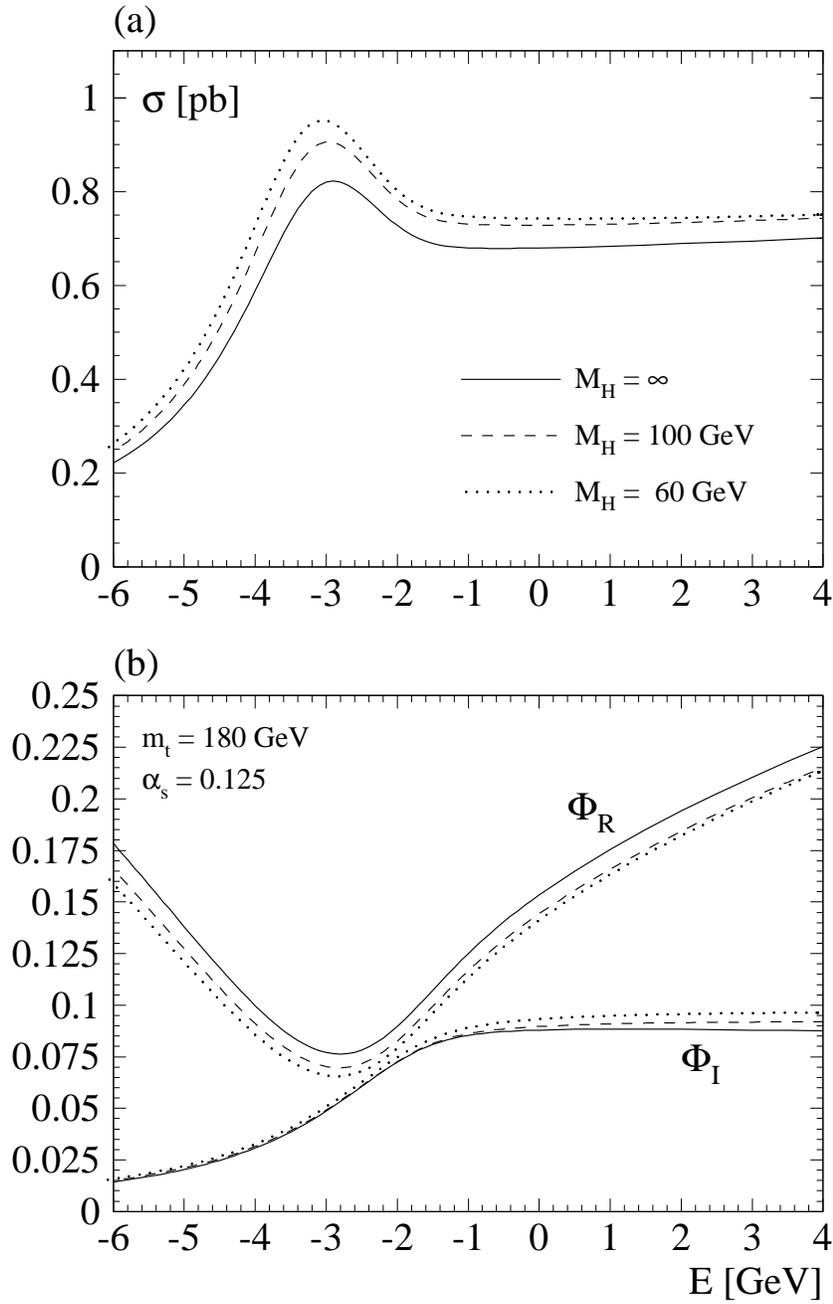}
\end{center}
\caption{\label{singlet}Higgs effects in $e^+e^- \rightarrow t\bar{t}$
	production for (a) the total cross section for unpolarized $e^\pm$
	and (b) the real ($\Phi_{\rm R}$) and imaginary ($\Phi_{\rm I}$)
	part of the function $\Phi(E)$ (see eq.(\ref{capphi})) for
	$M_H = 60$ GeV (dotted), $M_H = 100$ GeV (dashed) and
	$M_H \rightarrow \infty$ (solid).}
\end{figure}

\begin{figure}
\begin{center}
\leavevmode
\epsfxsize=11.5cm
\epsffile[90 120 470 700]{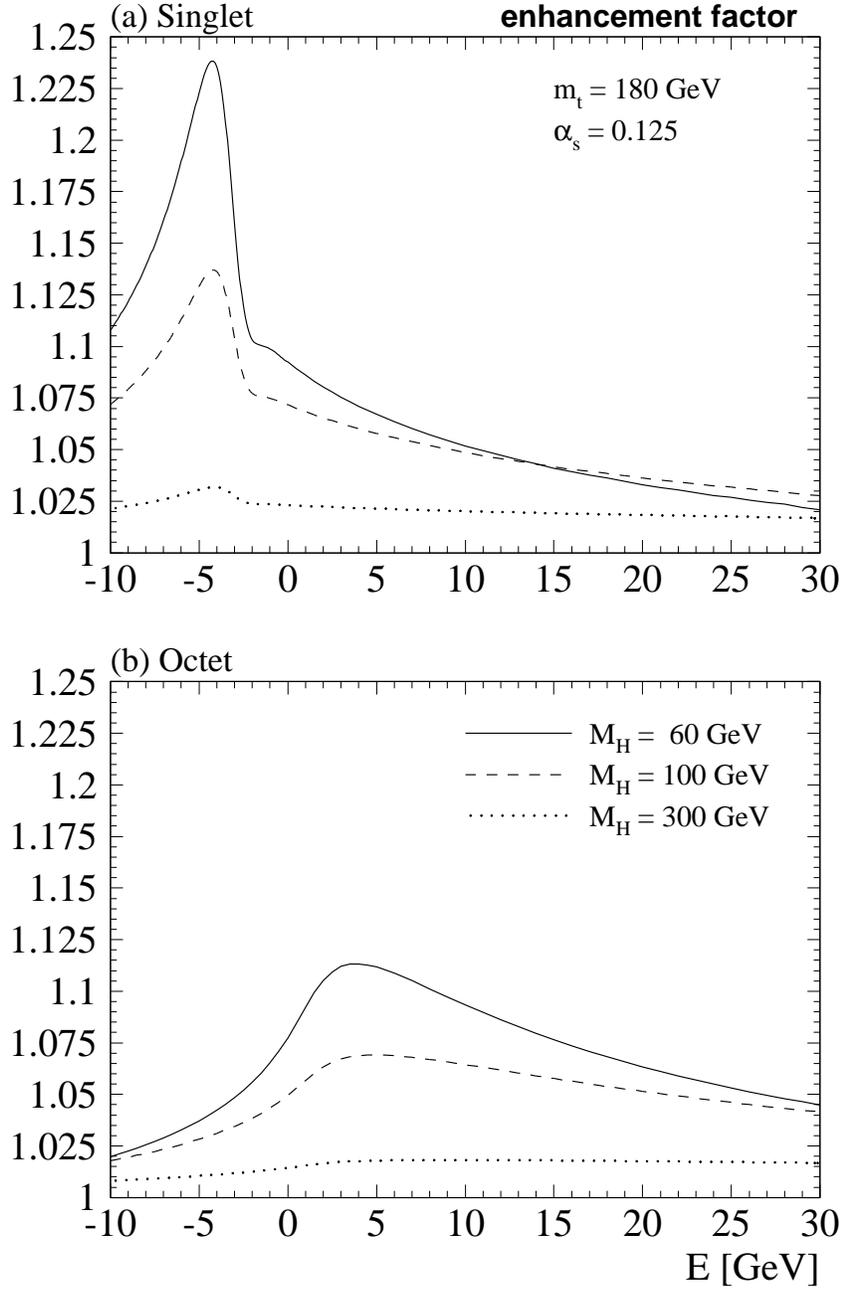}
\end{center}
\caption{\label{enhance}Enhancement factors due to exchange of a Higgs boson
	of mass $M_H = 60$ (solid), $100$ (dashed) and $300$ GeV (dotted)
	in \ttq--system: (a) singlet and (b) octet color configuration.}
\end{figure}

Considering first $e^+e^-$ annihilation, the \ttq--system is produced
in a color singlet state and the group theory factor is given by
$C_R = - 4/3$. Therefore, $V_{\rm QCD}$ is attractive enabling resonance
\ttq--production below threshold.
The width of the top quark largely smears these resonances such that in
fig.\ref{singlet}(a) only the $1s$--peak remains visible.
The solid line corresponds to the case without Higgs exchange.
The dashed one shows the behavior for $M_H = 100$ GeV, the dotted one for
$M_H = 60$ GeV.
In fig.\ref{enhance}(a) the ratio of the cross section for different
values of the Higgs mass to the one without Higgs is plotted in an energy
range from $-10$ to $+30$ GeV around threshold.
It strongly varies in the region around the $1s$--resonance for small values
of $M_H$: Reaching a maximum of about $24\%$, it rapidly goes down to
$10\%$ until the cross section passes into the continuum like region
built up by the overlapping higher resonances.
The maximum is less pronounced for a larger Higgs mass, flattening down to
about $3\%$ for $M_H = 300$ GeV. For $E > 0$ the energy dependence becomes
weaker with the curve getting nearly constant for $M_H = 300$ GeV.

In case of polarized electron and/or positron beams the angular distribution
and the components of the polarization vector are governed by the function
\beq{capphi}
\Phi(E) = {(1-4 \alpha_s/3 \pi) \over (1-8 \alpha_s/3 \pi)}\,
	{\int dp {p^3 \over m_t} F^*(\pb,E) G(\pb,E) \over
	 \int dp\, p^2 |G(\pb,E)|^2} \ \ .
\eeq
The corresponding formulae are given in \cite{hjkt,hjk}.
The function $\Phi(E)$ was first considered in \cite{sumino}.
Since the approach of \cite{jezk} has not been extended yet to the case of
$p$--wave, we include Higgs effects only through the effective potential
(\ref{v_eff}) when evaluating the function $\Phi(E)$.

Fig.\ref{singlet}(b) shows the real ($\Phi_{\rm R}$) and the imaginary
($\Phi_{\rm I}$) part of the function $\Phi(E)$
for two different values of the Higgs mass in comparison to the case of
a pure QCD--potential.

One can see that while the imaginary part grows with decreasing Higgs mass,
the real part gets smaller with the effect being stronger for the real part.
The dependence of $\Phi_{\rm I}$ on $M_H$ below threshold is almost negligible
while it gets comparable to that of $\Phi_{\rm R}$ for larger
energies. For $\Phi_{\rm R}$ it amounts about $10$--$20\%$ over the whole
range.

\begin{figure}
\begin{center}
\leavevmode
\epsfxsize=11.5cm
\epsffile[100 265 470 550]{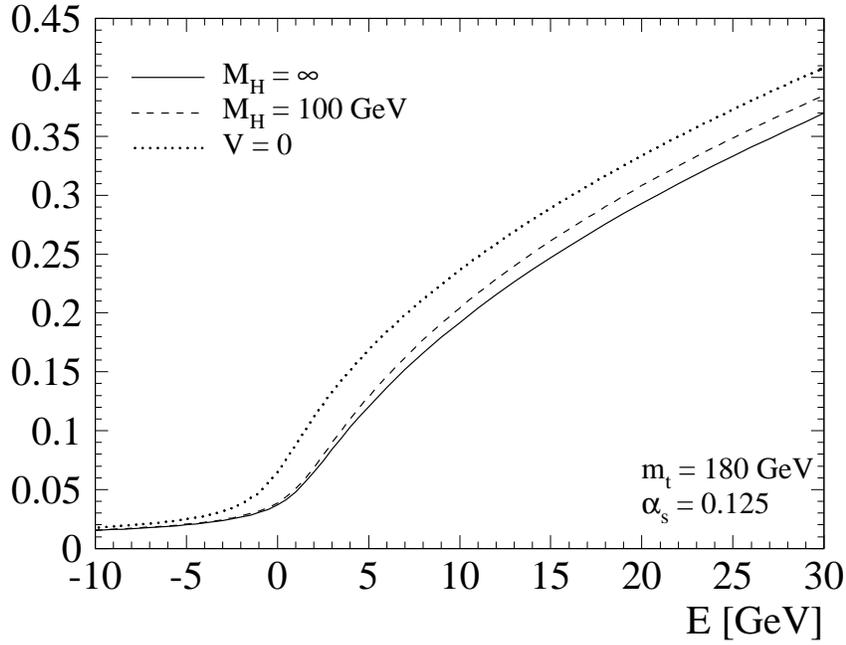}
\end{center}
\caption{\label{octet}
	Comparison of phase space factors $B(E)$ (\ref{phase_sp})
	for \ttq\ octet production,
	including QCD and Higgs effects for $M_H = \infty$ (solid line)
	and $M_H = 100$ GeV (dashed).
	The dotted line corresponds to production of free unstable quarks,
	cf.~eq.(\ref{phase_free}).}
\end{figure}

Turning to octet \ttq--production, the group factor becomes $C_R = 1/6$.
The effective potential is now repulsive, so that below threshold the
cross section essentially vanishes with a small smearing caused by the
non--zero width of the top quark.
Fig.\ref{octet} shows the threshold factor (\ref{phase_sp}) for the octet
case. Again the solid line corresponds to the case of a
pure QCD--potential and the dashed one to $V_{\rm eff}$ with $M_H = 100$~GeV.
The dotted line is now the phase space factor for free unstable particles
\beq{phase_free}
B^{(free)}(E) = \sqrt{\sqrt{E^2 + \Gamma_t^2} + E \over 2 m_t} \ \ ,
\eeq
evaluated by inserting the free Greens function
(\ref{prop_free}) into the right hand side of (\ref{phase_sp}).
For the pure QCD--potential the ``effective'' threshold is shifted towards
larger energies compared to the free case and the magnitude is lowered.
Higgs exchange cancels these effects to some extent.

As can be seen from fig.\ref{enhance}(b), the enhancement factor is
smoother than for the singlet case. It is also smaller with a maximum of
$11$--$12\%$ for $M_H = 60$ GeV, going down to less than $2\%$ for
$M_H = 300$ GeV, when the shape of the curve becomes almost totally flat.

\noindent
\sloppy
\raggedright

\end{document}